\documentclass[preprint,aps,amsmath,amssymb]{revtex4}
\usepackage{graphicx}
\usepackage{dcolumn}
\usepackage{color}

\usepackage[normalem]{ulem} 
\newcommand{\scrap}[1]{{\color{grey}{\sout{#1}}}}
\definecolor{darkgreen}{rgb}{0,0.5,0}
\definecolor{grey}{rgb}{.6,.6,.6}

\begin{document}

\title {Observation of the Kondo Screening Cloud of Micron Lengths}

\author {I. V. Borzenets$^{1}$$^{*}$, J. Shim$^{2}$, J. Chen$^{3}$, A. Ludwig$^{4}$,  A. Wieck$^{4}$, S. Tarucha$^{5}$, H.-S. Sim$^{2}$$^{*}$, and M. Yamamoto$^{5}$$^{*}$}

\affiliation{
$^{1}$Department of Physics, City University of Hong Kong, Kowloon Tong, Kowloon, Hong Kong
$^{2}$Department of Physics, Korea Advanced Institute of Science and Technology (KAIST), Daejeon 34141, South Korea
$^{3}$Department of Applied Physics, University of Tokyo, Bunkyo-ku, Tokyo, Japan
$^{4}$Faculty of Physics and Astronomy, Rurh-University Bochum, Bochum, Germany
$^{5}$Center for Emergent Matter Science (CEMS), RIKEN, Wako-shi, Saitama, Japan
$^{*}$Correspondence should be sent to I.V.B. (email: iborzene@cityu.edu.hk), H.-S.S (email: hssim@kaist.ac.kr), and M.Y. (email: michihisa.yamamoto@riken.jp) 
 }

\maketitle

\textbf{When a magnetic impurity exists in a metal, conduction electrons form a spin cloud that screens the impurity spin. This basic phenomenon is called the Kondo effect~\cite{Kondo_Review,Kondo_QD_4_Glazman}. Contrary to electric charge screening, the spin screening cloud\cite{Affleck2010,Gruner1974,Alternate_Proposal_1,Alternate_Proposal_2} occurs quantum coherently, forming spin-singlet entanglement with the impurity.
Although the spins interact locally around the impurity, the cloud can spread out over micrometers. The Kondo cloud has never been detected to date, and its existence, a fundamental aspect of the Kondo effect, remains as a long-standing controversial issue~\cite{No_Cloud_1_Boyce,Alternate_Proposal_5}. 
Here we present experimental evidence of a Kondo cloud extending over a length of micrometers comparable to the theoretical length $\xi_\mathrm{K}$.
In our device, a Kondo impurity is formed in a quantum dot (QD)~\cite{Kondo_QD_1_DGG, Kondo_QD_2_Leo,Kondo_QD_3_Leo, Kondo_QD_4_Glazman}, one-sided coupling to a quasi- one dimensional channel~\cite{Theory_Proposal_HS} that houses a Fabry-Perot (FP) interferometer of various gate-defined lengths $L > 1 \, \mu$m. 
When we sweep a voltage on the interferometer end gate separated from the QD by the length $L$
to induce FP oscillations in conductance, 
we observe oscillations in measured Kondo temperature $T_\mathrm{K}$, a sign of the cloud at distance $L$. 
For $L \lesssim \xi_\mathrm{K}$ the $T_\mathrm{K}$ oscillation amplitude becomes larger for the smaller $L$, obeying a scaling function of a single parameter $L/ \xi_\mathrm{K}$, while  for $L>\xi_\mathrm{K}$ the oscillation is much weaker. The result reveals that  $\xi_\mathrm{K}$ is the only length parameter associated with the Kondo effect, and that the cloud lies mostly inside the length $\xi_\mathrm{K}$ which reaches microns.
Our experimental method of using electron interferometers offers a way of detecting the spatial distribution of exotic non-Fermi liquids formed by multiple magnetic impurities or multiple screening channels~\cite{NonFermi_1_Cox, NonFermi_2_Affleck,Potok,Iftikhar} and solving long-standing issues of spin-correlated systems.
}

While Kondo physics for a single magnetic impurity has been well-established except for its spatial extension, our understanding of multiple impurity systems such as Kondo lattices, spin glasses, and high $T_c$ superconductors, is far from complete. In such systems, Kondo cloud length (or the spatial distribution of a Kondo cloud) with respect to the distance between impurities and other length parameters is crucial for understanding their properties. The detection and control of a Kondo cloud is therefore considered as one of ultimate goals of condensed matter physics.
There have been attempts to detect the Kondo cloud over 50 years~\cite{Affleck2010,Gruner1974,Alternate_Proposal_1,Alternate_Proposal_2,No_Cloud_1_Boyce,Alternate_Proposal_5,Theory_Proposal_HS,No_Cloud_3_Pruser,Alternate_Proposal_3,Alternate_Proposal_12,Alternate_Proposal_13,No_Cloud_5_Ivan,Alternate_Proposal_4,Alternate_Proposal_7,Alternate_Proposal_8}.
NMR (nuclear magnetic resonance) measurements have not found any signature of the cloud~\cite{No_Cloud_1_Boyce}.
STM  (scanning tunneling microscopy) experiments have shown a signature of the Kondo effect in a region away from a magnetic impurity, however, by distance much shorter than the cloud length~\cite{No_Cloud_3_Pruser}.
The difficulty lies in the fact that measuring spin correlation in the Kondo screening requires fast detection of tens of gigahertzs~\cite{Alternate_Proposal_3} or that there are complications from the atomic or electronic structure of samples. On the other hand, recent advancement of nanotechnology opens another way.
It allows one to prepare a single spin in a QD in contact with an electron reservoir, achieving systematic control of a single-channel Kondo state~\cite{Kondo_QD_1_DGG, Kondo_QD_2_Leo}.
The cloud length is typically $\xi_\mathrm{K} = \hbar v_\mathrm{F} / (k_\mathrm{B} T_\mathrm{K}) \sim 1 \, \mu$m for $T_\mathrm{K} \sim$ 1~K and the Fermi velocity $v_\mathrm{F} \sim 10^5$ m/s; a recent theoretical study~\cite{KondoEntanglement} of quantum entanglement shows that the Kondo state lies mostly within the distance $\xi_\mathrm{K}$ from the impurity, with a long algebraically decaying tail extending further.
Interesting proposals suggest using a finite-size electron reservoir, to observe competition between the cloud length and the reservoir size~\cite{Alternate_Proposal_12,Alternate_Proposal_13,No_Cloud_5_Ivan}.

\begin{figure}[h!]
\includegraphics[width=0.8\columnwidth]{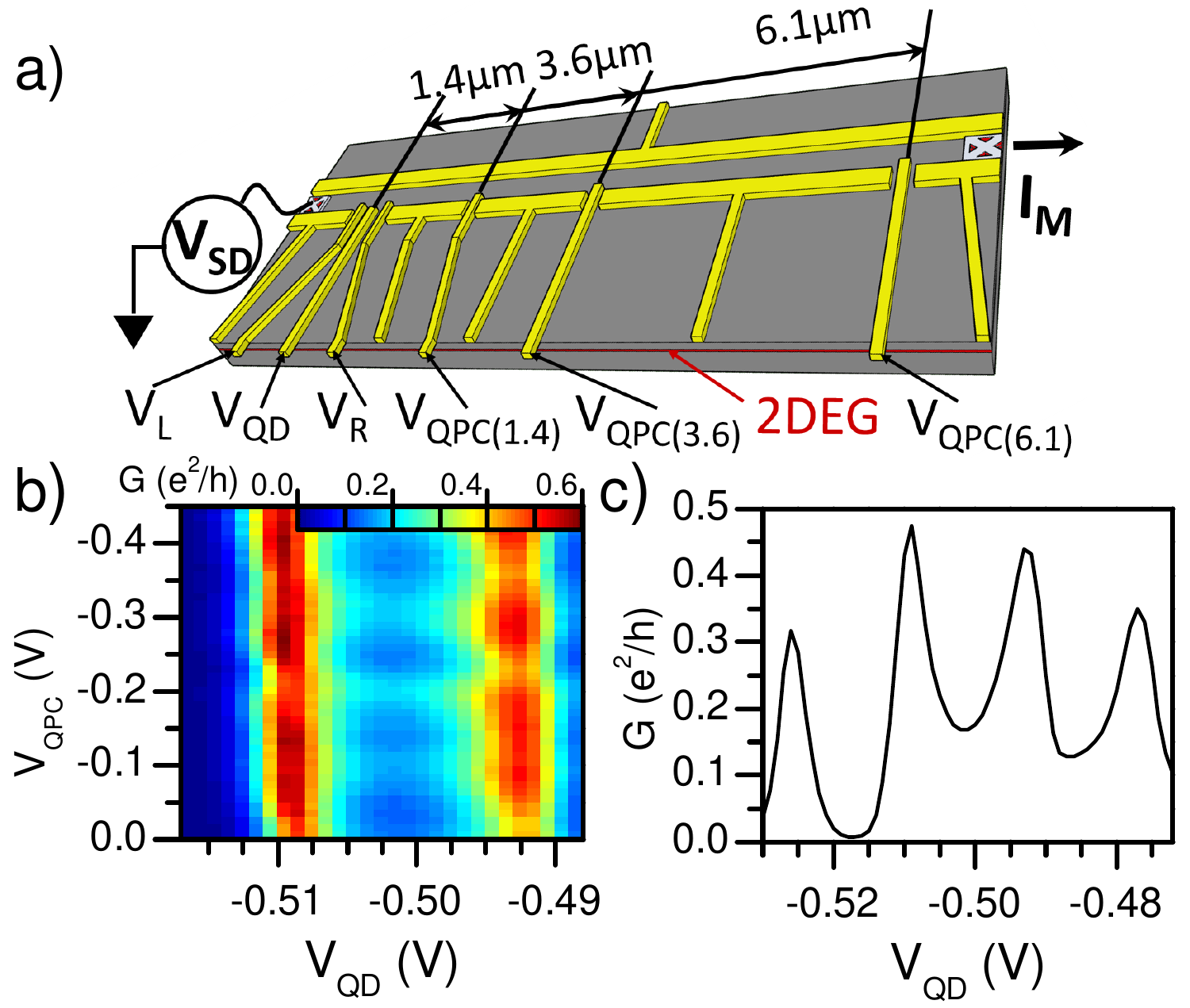}
\caption{\label{fig:overview} \textbf{Measurement setup and characterization.} a) Device and measurement schematic. The device consists of a quantum dot (QD) coupled to a 1D channel, in which three quantum point contact (QPC) gates embedded at distances $L$ = 1.4, 3.6, 6.1 $\mu$m from the QD. The activation of a QPC gate creates a Fabry-Perot (FP) cavity of the length $L$. 
The QD is tuned via a central plunger gate voltage $V_\mathrm{QD}$ and two side-gates $V_\mathrm{L}$ and $V_\mathrm{R}$. The device is measured via the lock-in method: a small AC voltage $V_\mathrm{SD}$ is applied, and the current $I_\mathrm{M}$ through the system is measured. b) Conductance $G$ of the device versus the plunger gate voltage $V_\mathrm{QD}$ and the $L=1.4 \, \mu m$ QPC gate voltage   $V_\mathrm{QPC}$. Coulomb blockade peaks are observed with respect to changing $V_\mathrm{QD}$. Oscillations associated with the FP cavity are seen with respect to changing $V_\mathrm{QPC}$. c) Conductance $G$ versus $V_\mathrm{QD}$ taken at $V_\mathrm{QPC}=0$.  Coulomb blockade as well as a region of enhanced conductance around $V_\mathrm{QD}=-0.50 \, \rm V$ associated with the Kondo valley are clearly observed. 
}
\end{figure}

In this work, instead of rigidly limiting reservoir size, we perturb the reservoir by inducing a weak barrier at a position $L$ far from the Kondo impurity and observe the resulting change in the Kondo effect with varying barrier position, following a recent proposal~\cite{Theory_Proposal_HS}.  Figure 1(a) shows the device and measurement schemes. An unpaired electron spin (magnetic impurity) is confined in a QD coupled to a one-dimensional (1D), long, ballistic channel~\cite{Takada, QD_1D}. The QD is in the Coulomb blockade regime. The 1D channel is tuned to contain several conducting channels and has three quantum point contact (QPC) gates placed away from the QD at lengths $L=1.4 \,\mu m$, $3.6 \, \mu m$, and $6.1\, \mu m$. Application of voltage  $V_\mathrm{QPC}$ to one of the QPC gates create a weak barrier so that a FP cavity~\cite{Fabry_Perot} of the length $L$ is formed between the QD and the QPC. The charging energy of the FP cavity is ineffective owing to strong coupling to the reservoir through multiple conducting channels. Changing $V_\mathrm{QPC}$ allows us to continuously tune the FP cavity between on- and off-resonances by altering the cavity length on the scale of Fermi wavelength $\Delta L\approx\lambda_\mathrm{F}$ ($\lambda_\mathrm{F} \sim 40 \, {\rm nm} \ll L, \xi_\mathrm{K}$) without affecting the potential profile around the QD. We find that changes in $V_\mathrm{QPC}$ strongly affect the measured Kondo temperature $T_\mathrm{K}$ when the cavity length $L$ is shorter than $\xi_{\mathrm{K} \infty} = \hbar v_\mathrm{F} / (k_\mathrm{B} T_{\mathrm{K} \infty})$. Here $\xi_{\mathrm{K}\infty}$ and $T_{\mathrm{K} \infty}$ are the bare theoretical cloud length and bare Kondo temperature defined in the absence of the QPCs or equivalently for the case of $L=\infty$. For $L\gg\xi_{\mathrm{K}\infty}$, changes of $V_\mathrm{QPC}$ have little effect on $T_\mathrm{K}$. This implies that the Kondo state extends over $\approx\xi_{\mathrm{K}\infty}$.

The device is defined using top gates deposited on top of a GaAs/AlGaAs 2DEG wafer, with electron mean free path of  about 8 $\mu$m (bigger than the device size). The QD population is controlled via a middle plunger gate voltage $V_\mathrm{QD}$. The coupling of the QD to the 1D channel on the right side and to the left side is adjusted by changing side gate voltages $V_\mathrm{R}$ and $V_\mathrm{L}$, respectively. The QD is  coupled stronger to the right 1D channel than the left channel, so that the Kondo state is sensitive to the FP cavity of the 1D channel. We tune $V_\mathrm{R}$ to change the bare cloud length $\xi_{\mathrm{K}\infty}$ and Kondo temperature $T_{\mathrm{K}\infty}$. Figure 1c shows conductance $G$ between the right 1D channel and the left lead via the QD as a function of the plunger gate voltage $V_\mathrm{QD}$ when the QPCs are turned off. Several Coulomb blockade peaks are clearly visible with the measured charging energy of $>500 \, \mu \mathrm{eV}$~\cite{TaruchaDot}. The Kondo effect is observed, manifesting itself in increased conductance in the valley region between two Coulomb blockade peaks~\cite{Kondo_QD_1_DGG, Kondo_QD_2_Leo,Kondo_QD_4_Glazman}.  The effect of the QPC gates is shown in Figure 1b which plots conductance $G$ versus  plunger gate voltage $V_\mathrm{QD}$ as well as the voltage $V_\mathrm{QPC}$ applied to the QPC gate at $L=1.4 \, \mu \mathrm{m}$. Both the Coulomb blockade peaks as well as the Kondo valley undergo FP oscillations with respect to changing $V_\mathrm{QPC}$; note that at small $V_\mathrm{QPC}$ (equivalent to the first few FP oscillations) there is no effect on the QD energy level. 
The resonance level spacing $\Delta \sim 300 \, \mu \mathrm{eV}$ estimated from the data is consistent with the cavity length $L=1.4 \, \mu \mathrm{m}$, considering the Fermi velocity $v_\mathrm{F}=2.46\times10^5$ m/s. The resonance level broadening $\sim 80 \, \mu \mathrm{eV}$ implies a weak barrier formed by $V_\mathrm{QPC}$ (Supporting Information).

\begin{figure}[h!]
\includegraphics[width=0.5 \columnwidth]{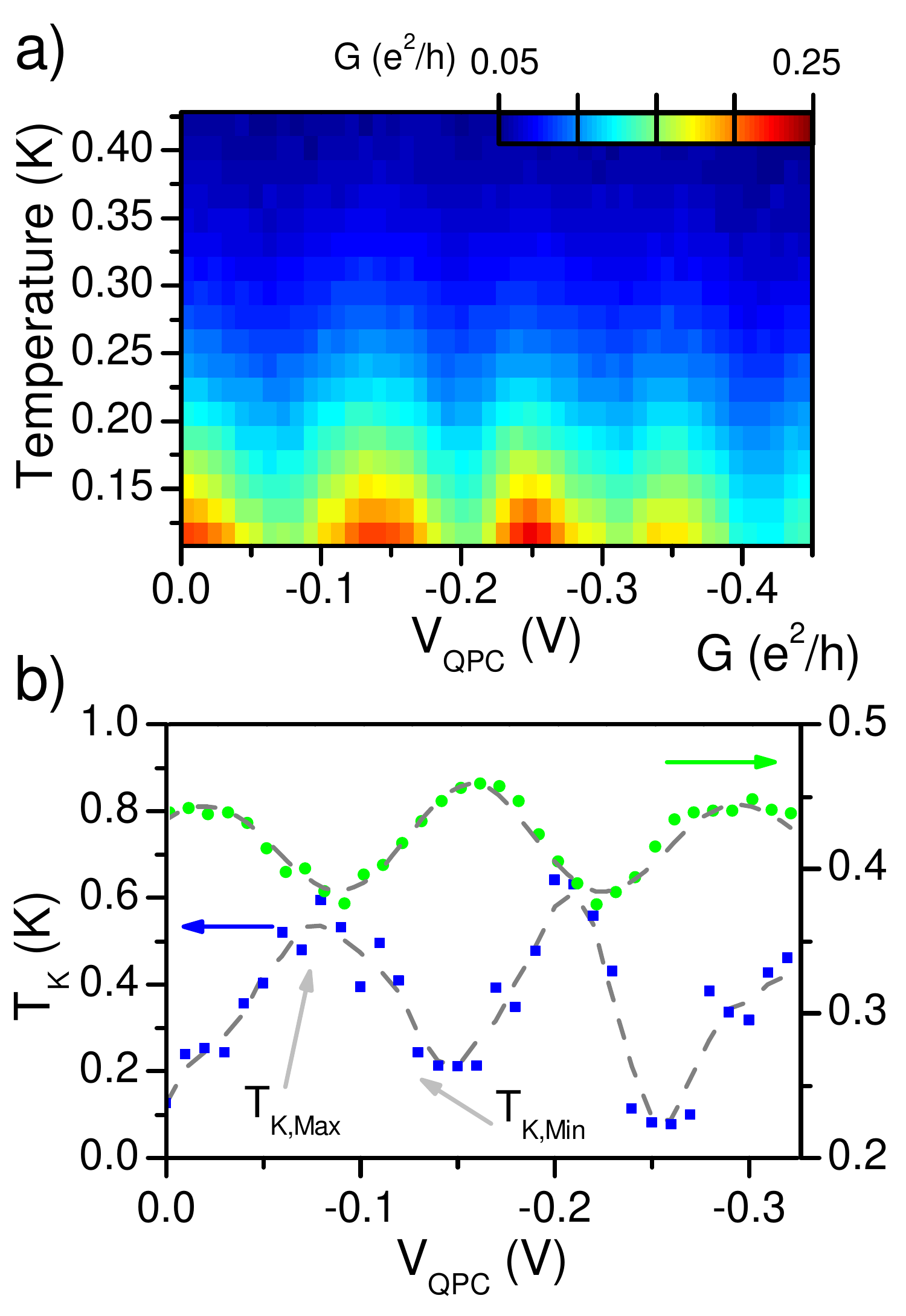}
\caption{\label{fig:overview} \textbf{Influence of Fabry-Perot interference on the Kondo effect. } a) Conductance $G$ at the center of the Kondo valley versus the QPC gate voltage $V_\mathrm{QPC}$ at $L=1.4 \, \mu \mathrm{m}$ and device temperature $T$. $G$ decreases with increasing $T$, indicative of the Kondo effect.  For each $V_\mathrm{QPC}$, the Kondo temperature $T_\mathrm{K}$ is extracted by fitting conductance versus  temperature to an empirical formula~\cite{Goldhaber-Gordon98}. b) Plot of the extracted Kondo temperature $T_\mathrm{K}$ versus $V_\mathrm{QPC}$ (blue), shown alongside the conductance $G$ (green). $T_\mathrm{K}$ oscillates with respect to $V_\mathrm{QPC}$, but in anti-phase with respect to conductance. The $T_\mathrm{K}$ oscillation amplitude is quantified by tracking the maximum $T_\mathrm{K,Max}$ and minimum $T_\mathrm{K,Min}$ of the first oscillation. 
}
\end{figure}

In order to see the effect of the FP oscillations on the Kondo state, we measure the conductance $G$ around the center of the Kondo valley as a function of the QPC gate voltage $V_\mathrm{QPC}$ at different temperatures $T$ (Figure 2a). The valley conductance decreases with increasing temperature, indicative of the Kondo state. For each value of $V_\mathrm{QPC}$ we extract a Kondo temperature $T_\mathrm{K}$ around the valley center by fitting conductance $G$ versus temperature $T$ to a well-known empirical formula (see Methods)~\cite{Goldhaber-Gordon98}. 
This method of extracting $T_\mathrm{K}$ is applicable for constant density of states of the reservoirs, but is still applicable to our QD coupled to the FP cavity, as the resonance level broadening is sufficiently large or the QPC barrier is weak (Supplementary Information). We find that the Kondo temperature $T_\mathrm{K}$ undergoes oscillations with respect to changing $V_\mathrm{QPC}$ (Figure 2b). Clearly, the Kondo state is affected by the perturbation at the location distant from the QD by microns. 
The oscillation shows that the electron density of the FP cavity at the Fermi level that is coupled with the Kondo impurity is different between on and off resonances. 
This implies that the Kondo coherence is extended through the entire FP cavity, supporting the picture of the spatial extension of the Kondo cloud. 
It also implies that the resonance level spacing $\Delta$ is larger than the Kondo temperature as theoretically expected.

The oscillations of  $T_\mathrm{K}$ are anti-phase with those of the conductance. This agrees with scattering theories combined with the Fermi liquid and numerical renormalization group (NRG) methods (Supplementary Information): In the on-resonance situations where the Fermi momentum $k_\textrm{F}$ satisfies $e^{i 2k_\textrm{F} L} = 1$, the FP cavity supports a maximum amount of electrons at the Fermi level, hence, the Kondo state with a maximum Kondo temperature is developed. The development effectively changes the resonance condition to off-resonance, resulting in a minimum value of the conductance, since the electrons gain twice the scattering phase shift $\pi/2$ off the Kondo impurity in the QD~\cite{pi_over_2}. The opposite happens in the off-resonance situations of $e^{i 2k_\textrm{F} L} = - 1$. Note that an evidence of the Kondo scattering phase $\pi/2$ was observed~\cite{Takada}, and  that the antiphase in Figure 2b can be considered as another evidence. In order to quantify the effect of the FP cavity on the Kondo state we track the maximum $T_\mathrm{K,Max}$ and minimum $T_\mathrm{K,Min}$ of the first oscillation of $T_\mathrm{K}$. Only the first oscillation is used as the subsequent oscillations require a higher $V_\mathrm{QPC}$ which will have a stronger effect on the QD due to capacitive coupling through the FP interferometer island.

\begin{figure}[hb!!]
\includegraphics[width=0.6 \columnwidth]{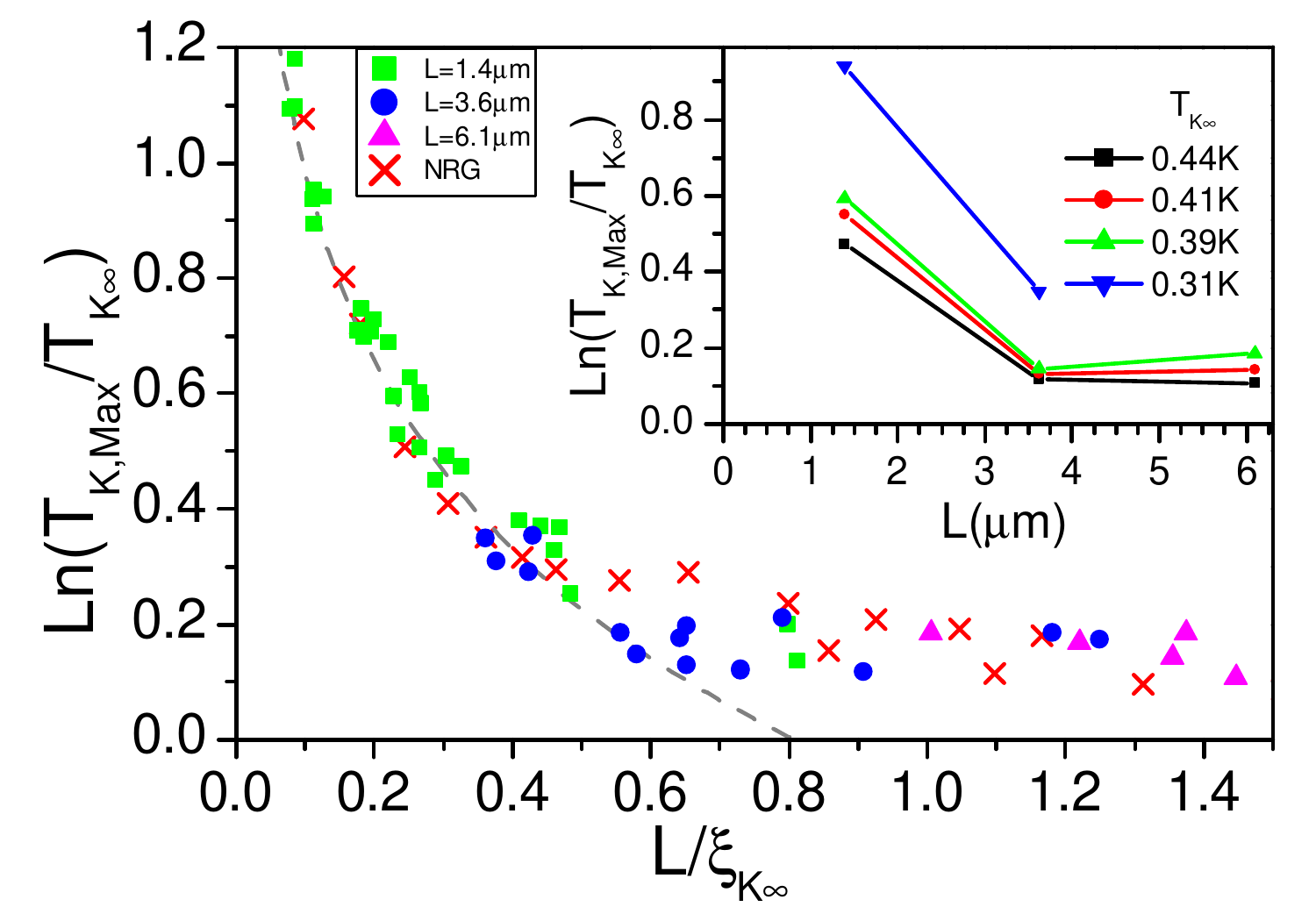}
\caption{\label{fig:overview} \textbf{Shape of the Kondo cloud revealed from modulation of the Kondo temperature.} Oscillation amplitude of Kondo temperature $T_\mathrm{K}$ as a function of the FP cavity length $L$. The amplitude is quantified by $\ln (T_\mathrm{K,Max}/T_{\mathrm{K}\infty})$, the maximum value $T_\mathrm{K,Max}$ of the oscillation normalized by the bare Kondo temperature $T_{\mathrm{K}\infty}$ in logarithmic scale, and $L$ is scaled by the bare Kondo cloud length $\xi_{\mathrm{K}\infty}$. 
The green squares are obtained from the oscillation of $T_\mathrm{K}$ with respect to changes in the voltage $V_\mathrm{QPC}$ of the QPC gate located at $L=1.4 \, \mu \mathrm{m}$ from the QD, the blue circles are for $L=3.6 \, \mu \mathrm{m}$, and the magenta triangles are for $L= 6.1 \, \mu \mathrm{m}$.
Different data points of the same symbol correspond to different $T_{\mathrm{K}\infty}$. The data are compared with theoretical results of NRG calculation (red crosses) and the scaling of  $\ln (T_\mathrm{K,Max} / T_{\mathrm{K} \infty}) = - \eta \ln (L / \xi_{\mathrm{K} \infty})$ with $\eta = 0.47$ (dashed curve).
Inset: The same data are shown as the oscillation amplitude of $T_\mathrm{K}$ versus the length $L$ not scaled by $T_{\mathrm{K} \infty}$. The different curves represent those having the same $T_{\mathrm{K}\infty}$. }

\end{figure}

We now discuss the dependence of the oscillation amplitude of $T_\mathrm{K}$ on the FP cavity length $L$ shown in the inset of Figure 3.
The amplitude is quantified by $\ln (T_\mathrm{K,Max}/T_{\mathrm{K}\infty})$, and we estimate the bare Kondo temperature from the oscillation as $T_{\mathrm{K}\infty}=(T_\mathrm{K,Max}$  $T_\mathrm{K,Min})^{\frac{1}{2}}$ (see Supplementary Information for the validity of the estimate); $T_{\mathrm{K}\infty}$ is not directly accessible as a QPC gate can form a barrier even when its voltage is turned off.
The oscillation amplitude is small, i.e., $T_\mathrm{K,max} \simeq 1.1 T_{\mathrm{K} \infty}$ or $T_\mathrm{K, max} \simeq 1.2 T_\mathrm{K, min}$, for the cavity length $L =  3.6 \, \mu \mathrm{m}$ and $6.1 \, \mu \mathrm{m}$ and $T_{\mathrm{K} \infty} =$ 0.39 - 0.44~K.
By contrast, the amplitude becomes drastically large for $L = 1.4 \, \mu \mathrm{m}$, for example, $T_\mathrm{K,max} \simeq 3.2 T_{\mathrm{K} \infty}$ or $T_\mathrm{K, max} \simeq 10 T_\mathrm{K, min}$ for $T_{\mathrm{K}\infty} = 0.31$~K. For each $L$, the amplitude becomes larger as $T_{\mathrm{K}\infty}$ is smaller. The result shows that the Kondo state is sensitive to the perturbation at distance $L \lesssim 3.6 \, \mu \mathrm{m}$ for $T_{\mathrm{K} \infty} =$ 0.39 - 0.44~K, while sensitive also at $L > 3.6 \, \mu \mathrm{m}$ for $T_{\mathrm{K} \infty} =$ 0.31~K. This implies that the cloud length is close to 3.6 $\mu \mathrm{m}$ for $T_{\mathrm{K}\infty} =$ 0.39 - 0.44~K, while larger than  3.6 $\mu \mathrm{m}$ for $T_{\mathrm{K}\infty} =$ 0.31~K. This finding is consistent with the bare cloud length $\xi_{\mathrm{K} \infty}$ estimated by the theoretical relation of  $\xi_{\mathrm{K} \infty} = \hbar v_\mathrm{F} / (k_\mathrm{B} T_{\mathrm{K} \infty})$, $\xi_{\mathrm{K} \infty} = 5.19$, 4.12, 3.92, and $3.65 \, \mu \mathrm{m}$ for $T_{\mathrm{K} \infty} = 0.31$, 0.39, 0.41, 0.44~K, respectively.

To see the universality of the results, we plot the oscillation amplitude of $T_\mathrm{K}$ versus the cavity length $L$ scaled by the bare cloud length $\xi_{\mathrm{K} \infty}$ in the main panel of Figure 3. Since we only analyze the first oscillation, transmission through the QPC is almost independent of $V_\mathrm{QPC}$ and common for all QPCs. We find that all data points fall onto a single curve as theoretically expected for a fixed transmission through the QPC. This scaling result is the evidence that $\xi_{\mathrm{K}}$ is the only length parameter associated with the Kondo effect. For $L \gtrsim  \xi_{\mathrm{K} \infty} $, the Kondo state is little affected by the perturbation at the distance $L$, as the maximum of the oscillation is 20\% larger than the minimum, $T_\mathrm{K, max} \simeq 1.2 T_\mathrm{K, min}$. 
As $L$ decreases below $\xi_{\mathrm{K} \infty}$, the oscillation amplitude becomes drastically larger, showing that  the maximum is 1000\% larger than the minimum ($T_\mathrm{K, max} \simeq 10 T_\mathrm{K, min}$) at $L \sim 0.1 \xi_{\mathrm{K} \infty}$. The increase follows the universal scaling~\cite{Theory_Proposal_HS} of $\ln (T_\mathrm{K,Max} / T_{\mathrm{K} \infty}) = - \eta \ln (L / \xi_{\mathrm{K} \infty})$ with a constant $\eta$ defined as modulation of the density of states set by the QPC pinch-off strength.
The plot is in good agreement with theoretical NRG calculation based on realistic parameters estimated from sample characterization (Supplementary Information).
The result is consistent with the theoretical result~\cite{KondoEntanglement} of the spatial distribution of the Kondo singlet entanglement that the main body of the Kondo cloud lies inside the length $\xi_{\mathrm{K}\infty}$ with a long tail extending beyond $\xi_{\mathrm{K} \infty}$. An equivalent, alternative picture is that for FP cavity of length $L$, there are  $L / \xi_{\textrm{K} \infty}$ $(\sim k_B T_{\textrm{K} \infty} / \Delta)$ localized single-particle states of size $\xi_{\textrm{K} \infty}$ in a row. When $L / \xi_{\textrm{K} \infty} > 1$,  the single-particle state located closest to the Kondo impurity forms the main body of the cloud and is wholly within $L$. Hence, the local perturbation at $L$ affects only the other single-particle states contributing to the cloud tail.
When $L / \xi_{\textrm{K} \infty} < 1$, the main body of the cloud extends beyond the distance $L$, hence, it is strongly affected by the perturbation.

Our result provides the evidence of the spatial distribution of the Kondo state over microns.
This gives insight into screening of a local impurity spin in a metal. It will be interesting to further study the spatial distribution, for example, engineering of the spatial spin screening and the entanglement of the Kondo state. For example, by applying large QPC gate voltage to our device, one can systematically study the screening cloud of a Kondo box~\cite{Alternate_Proposal_12,Alternate_Proposal_13,No_Cloud_5_Ivan}. It will be also valuable to study spin screening by multiple independent channels as in the multi-channel Kondo effects or in a situation of multiple impurities as in the two-impurity Kondo effects, as those effects are accompanied by non-Fermi liquids and quantum phase transition~\cite{NonFermi_1_Cox, NonFermi_2_Affleck}. Our strategy of detecting spin screening by applying a weak electrostatic gate at a position distant from an impurity spin is applicable to the realization~\cite{Potok,Iftikhar} of the effects with systematic control.

\section*{Methods}
\subsection*{Sample Preparation}
The system of the QD coupled to the quasi 1D FP resonant cavity was fabricated in a two-dimensional electron gas [2DEG, carrier density $n=3.12 \times 10^{11} \mathrm{cm}^{-2}$,  mobility $\mu=0.86 \times 10^6 \mathrm{cm}^2/\mathrm{V}\mathrm{s}$] heterointerface with standard surface Schottky gate technique.

\subsection*{Measurement}
 Measurements were performed in an Oxford Instruments MX100 dilution refrigerator with a base lattice temperature of $40 \, \mathrm{mK}$. The base electron temperature was measured to be $\sim 80 \, \mathrm{mK}$. (Electron temperature vs measured mixing chamber temperature was calibrated via analysis of the QD Coulomb peak width.)  Electron transport was measured using the lock-in method. AC voltage oscillation with a DC offset was applied to a sample via a divider with the AC excitation set to $3-15 \,  \mu \mathrm{V}$. The DC offset was set such as to achieve zero bias at the sample. (DC offset was varied in order to calculate the QD charging energy.) Current through the sample was measured using a home-made current sense amplifier with a current-sense resistor of $10 \, \mathrm{k}\Omega$ mounted on the mixing chamber. Gate voltages were controlled via a custom DAC.  Temperature of the device was changed globally via a heating coil at the mixing chamber. 

\subsection*{Kondo temperature estimation}
The Kondo temperature $T_\mathrm{K}$ is estimated by fitting the experimental data to the empirical formula~\cite{Goldhaber-Gordon98} of $G(T)=G_0(\frac{T_\mathrm{K}^{'2}}{T^2+T_\mathrm{K}^{'2}})^s$ with $T_\mathrm{K}^{'}=T_\mathrm{K}/ \sqrt{2^{1/s}-1}$, where the zero-temperature conductance $G_0$ and the exponent $s$ are fit parameters;
according to Ref.~\cite{Goldhaber-Gordon98} $s$ should not vary significantly from $0.22$; 
in this sense, the fitting has only two fit parameters $T_\mathrm{K}$ and $G_0$. This estimation works well when the density of states of the reservoirs coupled to a Kondo impurity is approximately energy independent near the Fermi level. In our setup in which the Kondo impurity in the QD is coupled to the FP cavity, the density of states is energy dependent and the estimation is applicable when the QPC barrier defining the FP cavity is so weak that the energy dependence is not significant. The applicability of the estimation is confirmed by our NRG calculation with the model parameters chosen from the experimental data (Supporting Information).
Note that the estimation of $T_\mathrm{K}$ directly from the experimental data is a merit of our experimental regime; it is unclear how to directly estimate $T_\mathrm{K}$ in the opposite regime where the QPC barrier is strong (corresponding to  a finite-size reservoir coupled to a QD~\cite{Alternate_Proposal_12,Alternate_Proposal_13,No_Cloud_5_Ivan}); it is because the temperature dependence of the conductance $G$ has a nontrivial feature due to the FP resonance when the QPC barrier is strong (the nontrivial feature is enhanced as the barrier becomes stronger, and the feature is not captured by the empirical formula), as shown in our NRG calculations.

\subsection*{Modeling}
For a model for our experiments, we theoretically study an Anderson impurity formed in a QD coupled to two 1D leads, one of which houses a FP cavity. The Hamiltonian of the model is found in the Supplementary Information. The parameters of the model are estimated from the experimental data. The QD spectral function is obtained by using the NRG method, and the temperature dependence of the conductance is computed~\cite{Kondo_Review} by combining the spectral function and the Fermi-Dirac distribution function.
The feature in Figure 2 that the oscillations of $T_\textrm{K}$ are anti-phase with those of the conductance is reproduced in our theory in two different ways. The feature is obtained by using the scattering-matrix formalism combined with the Fermi liquid theory and by taking into account of the scattering phase shift $\pi/2$ off the Kondo impurity. The anti-phase feature is also found in an independent way based on the NRG method and conductance calculation.
On the other hand, the NRG result of the Kondo temperature $T_\mathrm{K}$ in Figure 3 is obtained, in the same way with the experimental estimation of $T_K$, by fitting the conductance obtained by the NRG method to the empirical formula. The universal scaling behavior in Figure 3 is obtained by using the poorman scaling and it is confirmed by NRG calculations; the poorman scaling and the NRG give the same scaling form but with different values of the parameter $\eta$, as $\eta$ depends on the estimation method of $T_K$. The details are found in the Supplementary Information.

\begin{acknowledgments}  I.V.B. acknowledges CityU New Research Initiatives/Infrastructure Support from Central (APRC): 9610395, and the Hong Kong Research Grants Council (ECS) Project: 9048125. S.T. and M. Y. acknowledge KAKENHI (GrantNo. 38000131). M.Y. acknowledges KAKENHI (GrantNo.18H04284) and CREST-JST (No. 18071350). H.-S.S. acknowledges support by Korea NRF via the SRC Center for Quantum Coherence in Condensed Matter (GrantNo. 2016R1A5A1008184).

\end{acknowledgments}



\pagebreak
\begin{center}
\textbf{\large Supplementary Information: Observation of the Kondo Screening Cloud of Micron Lengths}
\end{center}

\newcommand{\beginsup}{%
        \setcounter{equation}{0}
        \renewcommand{\theequation}{S\arabic{equation}}%
        \setcounter{table}{0}
        \renewcommand{\thetable}{S\arabic{table}}%
        \setcounter{figure}{0}
        \renewcommand{\thefigure}{S\arabic{figure}}%
     }

\beginsup

\section{Calibration of Measured Sample Temperature}
For temperatures below $\sim 500 \, \mathrm{mK}$ it becomes increasingly difficult to thermalize hot electrons sent to the sample from the measurement electronics. In this measurement setup, the hot electrons were cooled at the mixing chamber via a thermalization coil as well \scrap{,} as via copper powder filters. Never the less, calibration was required to match the measured mixing chamber lattice temperature to the actual electron temperature in the sample. This was achieved by measuring the shape of the Coulomb peaks of the QD with all the QPCs turned off (Figure S1c). Picking a peak that does not feature any Kondo temperature, we expect that the conductance will be proportional to: $G(\delta V_\mathrm{G})\propto \cosh^{-2}(\frac {(a\delta V_\mathrm{G})}{2k_\mathrm{B}T})$\cite{TaruchaDot_S}. Here, $\delta V_\mathrm{G}$ is the shift in plunger gate voltage away from the Coulomb peak center, $k_\mathrm{B}$ is the Boltzmann constant. The constant $a$ is independent of electron temperature and should only depend on the sample geometry.  (Figure S1a shows the fit of Conductance $G$ versus gate voltage $V_\mathrm{G}$ taken at a measured temperature of $T_\mathrm{Measured}=300 \, \mathrm{mK}$). Indeed, above $T_\mathrm{Measured}>600 \, \mathrm{mK}$ we see that $a$ has little to no variation, therefore, the mixing chamber thermometer temperature and the electron temperature are the same $T_\mathrm{Calibrated}=T_\mathrm{Measured}$. By looking at how the constant $a$ evolves for $T_\mathrm{Measured}<600 \, \mathrm{mK}$ we can now build the calibration for the electron temperature $T_\mathrm{Calibrated}=F(T_\mathrm{Measured})$ as shown in Figure S1b.  Throughout the text, the temperature $T$ referers to the calibrated temperature $T_\mathrm{Calibrated}$.

\begin{figure}[h]
\includegraphics[width=0.5 \columnwidth]{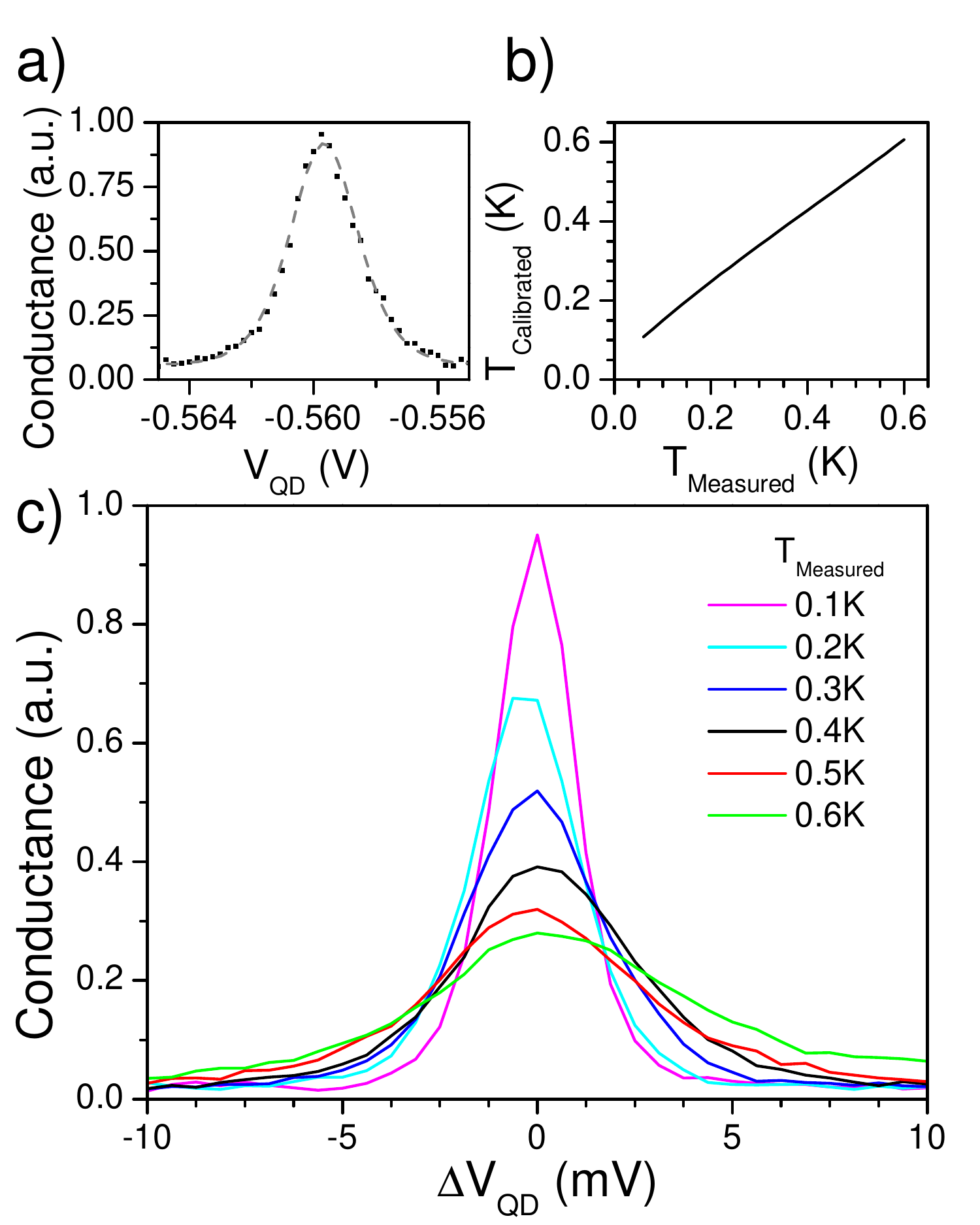}
\caption{\label{fig:overview} a) Conductance versus QD gate voltage $V_\mathrm{G}$  taken at a Temperature of $300 \, \mathrm{mK}$. The gray lines show the fit of the data to theoretical lineshape for a Coulomb Blockade peak.  b) The electron temperature in our device $T_\mathrm{Calibrated}$ versus the lattice temperature measured by a thermometer located at the mixing chamber $T_\mathrm{Measured}$. c) Conductance of the Coulomb blockade peaks vs gate voltage (around the peak center) $\delta V_\mathrm{G}$ shown for several values of the measured temperature.   
}
\end{figure}

\section{QPC Barrier Strengths}
The strengths of the oscillations in the Kondo Temperature $T_\mathrm{K,Max}/T_{\mathrm{K}\infty}$ depends on the QPC pinch-off strength $\alpha$. Here we define $\alpha=1- (t_0/t)^2$ with $t_0$ being the hopping energy across the QPC.  ($\alpha=0$ means there is no pinchoff present due to the QPC, and $\alpha=1$ means that the QPC fully decouples the FP cavity from the rest of the wire. See the Hamiltonian in Eq.~\eqref{rightwire}.) 
Below we estimate $\alpha$ in three different ways, which show $\alpha \sim 0.1$.
Fitting the main results (main text Figure 3) to the NRG calculations, we arrive at an $\alpha \sim 0.1$. In addition, we find that $\alpha$ can be obtained from the ratio of on- and off-resonance conductances of the Kondo valley at $T=0$: $\alpha=1-(G_\mathrm{0,Min}/G_\mathrm{0,Max})^{0.5}$. The conductances at zero temperature can be extracted in the similar method to $T_\mathrm{K}$ by fitting to the empirical formula~\cite{Empirical_Kondo} (The fitting will be discussed below in details). Averaging the data over several instances $V_\mathrm{L}$ and $V_\mathrm{R}$ we arrive at an $\alpha\sim 0.1$. Finally we estimate $\alpha$ also by considering the behavior of the system away from the Kondo regime. The c strength is proportional to the fluctuations in the local carrier density $\rho$ inside the FP cavity: $\alpha=1-(\rho_\mathrm{Min}/\rho_\mathrm{Max})^{0.5}$, where $\rho_\mathrm{Max(Min)}$ is the maximum (minimum) value of the fluctuations in $\rho$. In turn, the effective coupling strength $\Gamma_\mathrm{r}$ of the QD to the FP cavity is proportional to the local carrier density $\rho$. The coupling  strength $\Gamma_\mathrm{r}$ directly affects the width of the Coulomb blockade peak with respect to gate voltage $V_\mathrm{G}$. Indeed, as we tune the FP resonance by changing the QPC gate voltage $V_\mathrm{QPC}$, the Coulomb blockade peak undergoes fluctuations in width in synchronisity with fluctuations in conductance (Figure S3).  Taking the first oscillation we arrive at the QPC pinchoff strength $\alpha=1-(\mathrm{Width}_\mathrm{Min}/\mathrm{Width}_\mathrm{Max})^{0.5} \approx 0.1$.

\begin{figure}[h]
\includegraphics[width=0.5 \columnwidth]{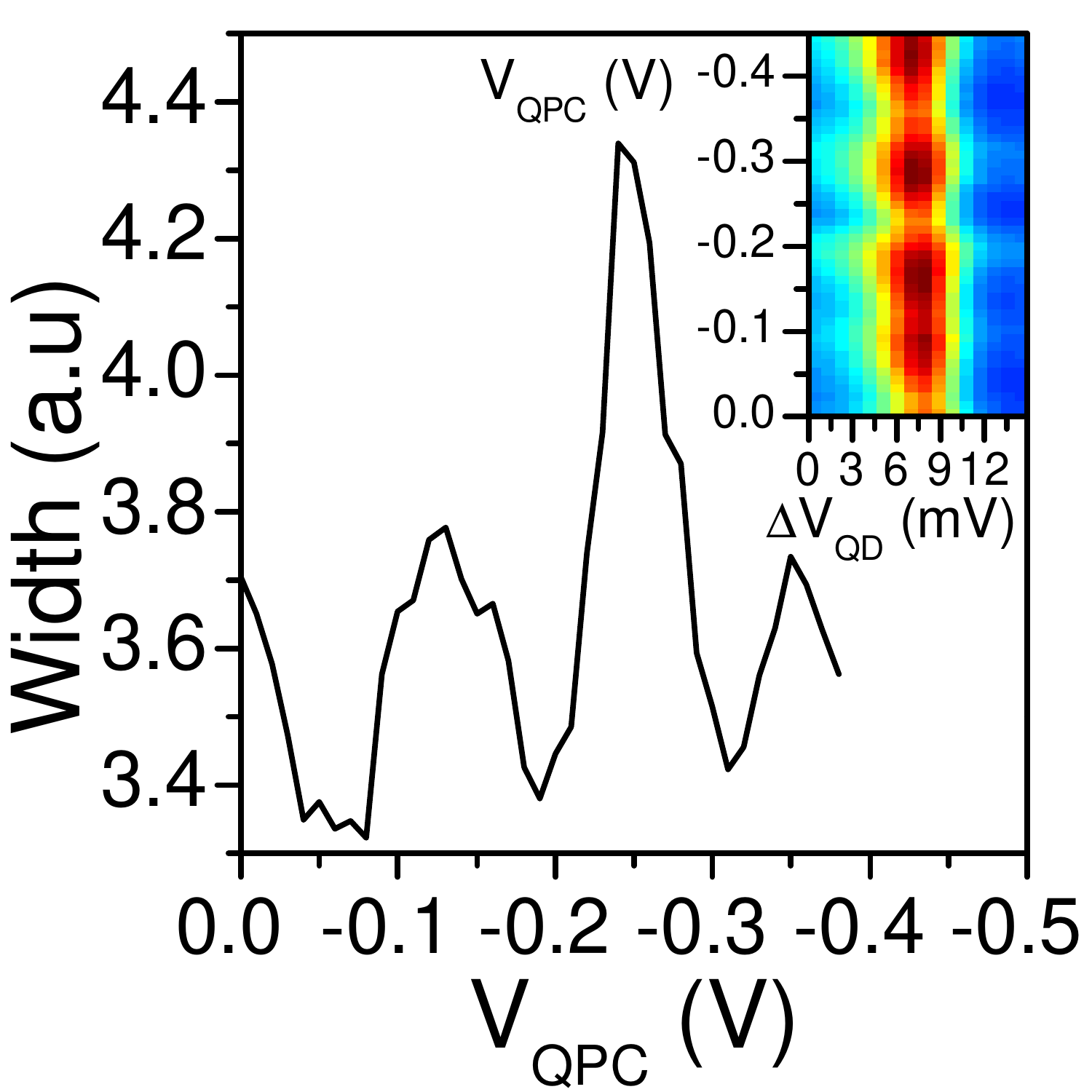}
\caption{\label{fig:overview}  Full width at half maximum of a Coulomb blockade peak versus QPC gate voltage $V_\mathrm{QPC}$. Data are shown for a Coulomb peak away from the Kondo valley. Inset) Conductance $G$ through the QD as a function of (shifted) gate voltage $\delta V_\mathrm{G}$ and QPC gate voltage $V_\mathrm{QPC}$. Ocillations of both peak conductance and Coulomb peak width with respect to $V_\mathrm{QPC}$ are clearly observed.  
}
\end{figure}

\section{Theoretical model}\label{t10}

The essential features of the experimental data are in good agreement with the numerical renormalization group (NRG) calculation of a model described below.

\subsection{Model Hamiltonian}\label{t11}

The experimental setup can be described by a theoretical model based on the Hamiltonian $H$,
\begin{align}\label{modelH}
H = H_\mathrm{QD} + H_\mathrm{l} + H_\mathrm{r} + H_\mathrm{tun}.
\end{align}
In the model, the quantum dot (QD) is described by an Anderson impurity, $H_\mathrm{QD} = \sum_{\sigma=\uparrow, \downarrow} \epsilon_\mathrm{QD} d_\sigma^\dagger d_\sigma + U d_\uparrow^\dagger d_\uparrow d_\downarrow^\dagger d_\downarrow$, where $d^\dagger_\sigma$ is the operator creating an electron with spin $\sigma$ in the QD, $\epsilon_\mathrm{QD}$ is the onsite energy of the electron, and $U$ is the charging energy. The QD couples with the two wires by electron tunneling. The tunneling Hamiltonian is $H_\mathrm{tun} = \sum_{w=\mathrm{l}, \mathrm{r}} \sum_{\sigma=\uparrow, \downarrow} t_w c_{w 1 \sigma}^\dagger d_\sigma + \text{h.c.}$, where $t_w$ is the tunneling amplitude between the QD and the first site $j=1$ of the wire $w = \mathrm{l}, \mathrm{r}$, $c_{w j \sigma}^\dagger$ is the operator creating an electron with spin $\sigma$ in the site $j$ of the wire $w$, and h.c. stands for hermitian conjugate.
The left wire is described by a semi-infinite tight-binding chain,  $H_\mathrm{l} = - \sum_{\sigma = \uparrow, \downarrow} \sum_{j=1}^\infty t c_{\mathrm{l} j\sigma}^\dagger c_{\mathrm{l} (j+1)\sigma} + \text{h.c.}$ $t$ is the electron hopping amplitude between neighboring sites of the wire. The right wire is also described by a semi-infinite chain,
\begin{align}
H_\mathrm{r} = & - \sum_{\sigma = \uparrow, \downarrow} [ \sum_{j=1}^{L-1} t c_{\mathrm{r} j\sigma}^\dagger c_{\mathrm{r} (j+1)\sigma} + \sum_{j=L+1}^{\infty}   t c_{\mathrm{r} j\sigma}^\dagger c_{\mathrm{r} (j+1)\sigma} \nonumber \\
& +  t_0 c_{\mathrm{r} L \sigma}^\dagger c_{\mathrm{r} (L+1) \sigma} + \text{h.c.} ].
\label{rightwire}
\end{align}
In the right wire, the electron hopping amplitude $t_0$ between the sites $j=L$ and $j=L+1$ is smaller than the hopping amplitude $t$ of the other part. This describes the effect of the quantum point contact (QPC) formed by $V_\mathrm{QPC}$ in the right wire in the experimental setup. As a result, the FP resonant cavity is formed between the first site $j=1$ and the site $j=L$.

\subsection{Fabry-Perot resonant cavity}\label{t12}

The hybridization function $\Gamma_\mathrm{r}$ between the right wire and the QD depends on whether a resonant level of the FP cavity is aligned with the Fermi level~\cite{Park13}.
When the Fermi level $\epsilon_\mathrm{F}$ is located at the center of the energy band 
(i.e.,  $\epsilon_\mathrm{F} = 0$),
we compute $\Gamma_\mathrm{r}(E)$ at energy $E$ near the Fermi level, using the Hamiltonian in Eq.~\eqref{modelH},
\begin{align}\label{Gamma_r}
\Gamma_\mathrm{r}(E) \approx \frac{\Gamma_{\mathrm{r} \infty}}{(1-\alpha)\text{cos}^2(\frac{\pi E}{\Delta} + k_\mathrm{F} L) + \frac{1}{1-\alpha}\text{sin}^2(\frac{\pi E}{\Delta} + k_\mathrm{F} L)}.
\end{align}
$\Gamma_{\mathrm{r} \infty} = \frac{t_\mathrm{r}^2}{2t^2} \sqrt{4t^2 - E^2}$ is the hybridization function in the absence of the QPC (i.e., in the case of  $t_0 = t$),
$\Delta = \pi \hbar v_\mathrm{F} / L$ is the resonance level spacing of the cavity, and $v_\mathrm{F}$ is the Fermi velocity.
$\alpha = 1 - (t_0/t)^2 \in [0, 1]$ parameterizes the energy broadening of the resonance level by the QPC; larger broadening occurs for smaller $\alpha$.
$k_\mathrm{F} L$ describes the resonance condition at the Fermi wavevector $k_\mathrm{F}$. In the situation of on-resonance (when the center of a resonance level is aligned with the Fermi level), $e^{2ik_\mathrm{F}L} = 1$. In the off-resonance case, $e^{2ik_\mathrm{F}L} = -1$.
 
On the other hand, the left wire does not have a FP cavity. The hybridization function $\Gamma_\mathrm{l}$ between the left wire and the QD is found as $\Gamma_\mathrm{l} (E) =  \frac{t_\mathrm{l}^2}{2t^2} \sqrt{4t^2 - E^2}$.

\subsection{Parameter estimation}\label{t13}

We estimate the parameters of the model from the experimental data.
The estimation is shown in Table~\ref{par}.

\begin{center}
\begin{table}[b]
\begin{tabular}{ccc}
\hline\hline
parameter & $\quad$ & value
\\
\hline
$U$ & $\quad$ & 600 $\mu$eV
\\
$\epsilon_\mathrm{QD}$ & $\quad$ & 300 $\mu$eV ($= -U/2$)
\\
$t_\mathrm{l}$ & $\quad$ & 150 $\mu$eV
\\
$t_\mathrm{r}$ & $\quad$ & 547.5 $\mu$eV to 600 $\mu$eV ($3.65 t_\mathrm{l}$ to $4 t_\mathrm{l}$)
\\
$t$ & $\quad$ & 3 meV
\\
$\Delta$ & $\quad$ & 300 $\mu$eV,  120 $\mu$eV, 75 $\mu$eV
\\
$\alpha$ & $\quad$ & $0.1$
\\
$T$ & $\quad$ & $0.13$ K to $0.5$ K
\\
\hline\hline
\end{tabular}
\caption{Parameter estimation used in the analysis.}
\label{par}
\end{table}
\end{center}

The QD charging energy $U$ is chosen as 600 $\mu$eV, based on the observed Coulomb diamond. The QD onsite energy is chosen as $\epsilon_\mathrm{QD} = -U/2$, since the QD is in the Kondo valley in the experiment. 
The estimated values of the tunneling amplitudes $t_\mathrm{l}$ and $t_\mathrm{r}$ agree with the experimental regime of $t_\mathrm{l} \ll t_\mathrm{r}$, and with the observed QD conductance ranging from  $0.4e^2/h$ to $0.6e^2/h$. Following the experiment, $t_\mathrm{l}$ is fixed and $t_\mathrm{r}$ is tuned in the theoretical analysis.
The hopping energy $t$ of the wires is sufficiently larger than the other energy scales.
The resonance level spacing  $\Delta$ is estimated, based on the observed resonant conductance peaks. We obtain three different values of $\Delta$, corresponding to the three cavity lengths $L$ of the experiment.
The resonance level broadening parameter $\alpha \in [0,1]$ is chosen as $\alpha = 0.1$ with which the experimental data are well described by the model in Eq.~\eqref{modelH}. The small value of $\alpha = 0.1$ agrees with the experimental situation that the QPC gate voltage is weak (hence the resonant conductance peaks have wide width).

\subsection{NRG calculation}\label{t14}

Using the Hamiltonian in Eq.~\eqref{modelH} and the NRG~\cite{Wilson75, Bulla08}, we compute the conductance $G$ between the left and right wires of the experimental setup. In the computation, the NRG discretization parameter is $\Lambda = 2$, $500$ states are kept at each iteration, and $z$-averaging is done for $z=0, 1/n_z, 2/n_z, \dots, (n_z - 1)/n_z$ with $n_z=8$.

Using the NRG result of the QD spectral function $A_{\mathrm{QD} \sigma}(E, T)$ of spin $\sigma$ at energy $E$, we compute the temperature $T$ dependence of the conductance~\cite{Hewson93, Pustilnik04},
\begin{align}\label{G}
G(T) = \frac{e^2}{h} \sum_{\sigma=\uparrow, \downarrow} \int dE \left(-\frac{\partial f}{\partial E}\right) \frac{4 \pi \Gamma_\mathrm{l} \Gamma_\mathrm{r}}{\Gamma_\mathrm{l} + \Gamma_\mathrm{r}} A_{\mathrm{QD} \sigma},
\end{align}
where $f$ is the Fermi-Dirac distribution.

To compare the experimental data with the NRG result in Fig.3 of the main text, we apply the same method for the estimation of the Kondo temperature to the experimental data and to the NRG result. The method is the fitting to the empirical formula~\cite{Goldhaber-Gordon98} 
\begin{equation} G(T)=G_0(\frac{T_\mathrm{K}^{'2}}{T^2+T_\mathrm{K}^{'2}})^s, \quad \quad \quad T_\mathrm{K}^{'}=T_\mathrm{K} \frac{1}{\sqrt{2^{1/s}-1}} \label{empiricalF} \end{equation} in the temperature window from $T=0.13$ K to  0.5 K (the window available in the experiment).
Here $G_0$ is the zero temperature conductance and $s \approx 0.2$ is a fit parameter.
The applicability of the empirical formula to our case of the QD coupled to the FP cavity is discussed in Sec.~\ref{t32}.

\begin{figure}[t]
\centerline{\includegraphics[width=0.5\textwidth]{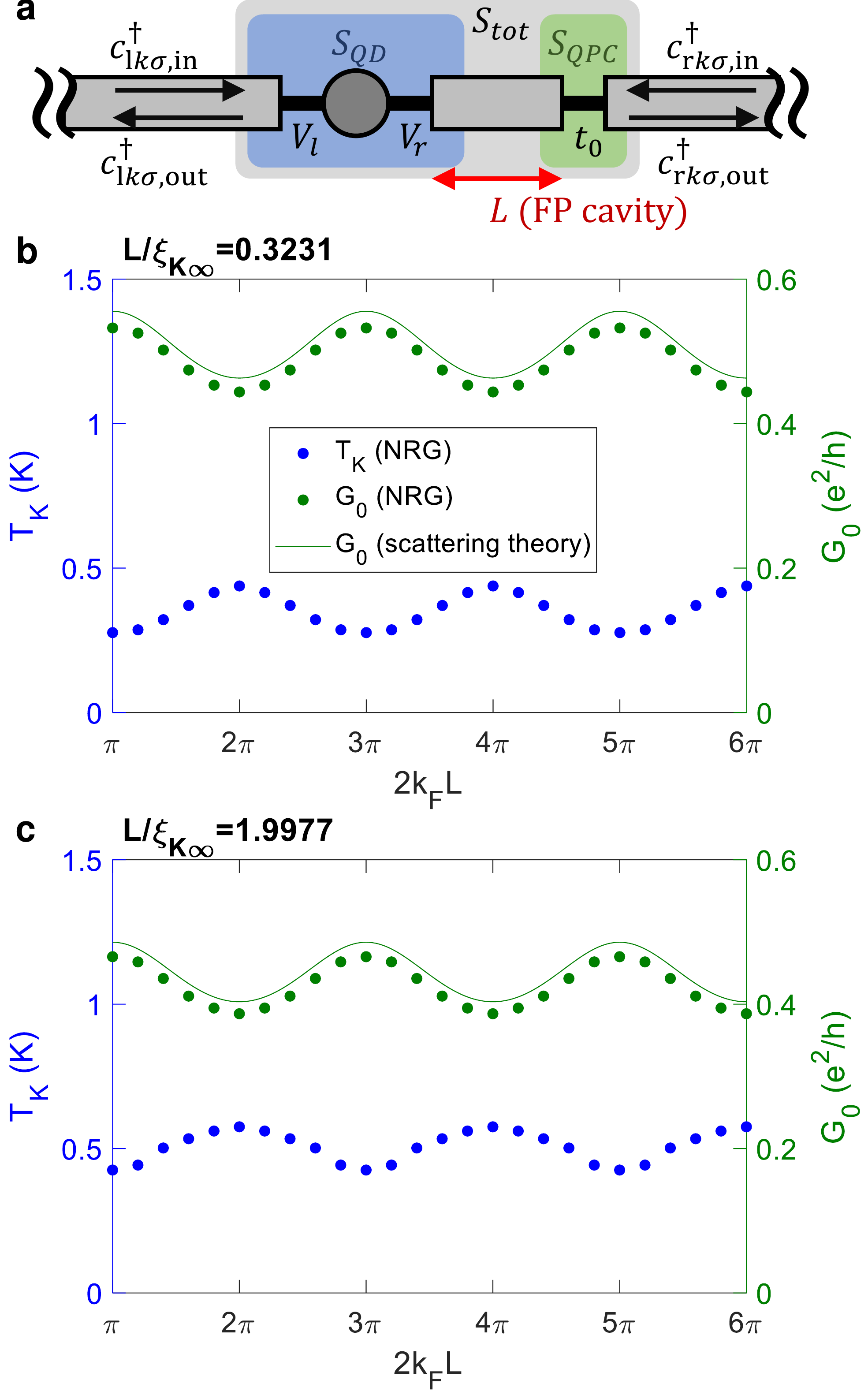}}
\caption{a) Schematic view for scattering matrices. b-c) The oscillations of the zero-temperature conductance $G_0$ and the Kondo temperature $T_\mathrm{K}$ as a function of $2 k_\textrm{F} L$ for (b) $L / \xi_{\mathrm{K} \infty} = 0.3231$ and (c) 1.9977. The two oscillations are out of phase by $\pi$ in both the cases of $L / \xi_{\mathrm{K} \infty}$. $G_0$ is obtained by using the scattering theory in Eq.~\eqref{T_tot} (see blue solid curves) and also by the NRG calculation (blue circles), while $T_\mathrm{K}$ is obtained by applying the empirical formula to the temperature window of [0 K, 0.5 K].
}
\label{fig_t1}
\end{figure}

\section{Conductance oscillation}\label{t20}

To explain the conductance oscillation in Fig. 2 of the main text, we construct a scattering theory based on the Fermi liquid theory. The result of the scattering theory agrees with the experimental data and the NRG result.

\subsection{Scattering theory}\label{t23}

We compute the scattering matrix $S_\mathrm{tot}$ of the region combining the QD, the FP cavity, and the QPC,
\begin{align}\label{S_tot}
\left[
\begin{array}{c}
c^\dagger_{\mathrm{l}k \sigma,\mathrm{out}}
\\
c^\dagger_{\mathrm{r}k \sigma,\mathrm{out}}
\end{array}
\right]
= S_\mathrm{tot}
\left[
\begin{array}{c}
c^\dagger_{\mathrm{l}k \sigma,\mathrm{in}}
\\
c^\dagger_{\mathrm{r}k \sigma,\mathrm{in}}
\end{array}
\right]
= \left[
\begin{array}{cc}
r_\mathrm{tot} & t'_\mathrm{tot}
\\
 t_\mathrm{tot} & r'_\mathrm{tot}
\end{array}
\right]
\left[
\begin{array}{c}
c^\dagger_{\mathrm{l}k \sigma,\mathrm{in}}
\\
c^\dagger_{\mathrm{r}k \sigma,\mathrm{in}}
\end{array}
\right],
\end{align}
where $c^\dagger_{w k \sigma,\mathrm{in (out)}}$ creates an electron with momentum $k$ and spin $\sigma$ incoming from the wire $w$ to the scattering region (outgoing from the scattering region to the wire $w$).  See Fig.~\ref{fig_t1}a. The scattering matrix formalism is applicable to the low-temperature Fermi-liquid regimes. At zero temperature, the electron conductance $G$ between the left and right wires is written as $G(T=0) = G_0 = 2e^2 |t_\mathrm{tot}|^2 / h$~\cite{Pustilnik04}.

$S_\mathrm{tot}$ is obtained by combining the scattering matrix $S_\mathrm{QD}$ of the QD, the scattering matrix  $S_\mathrm{QPC}$ of the QPC, and the dynamical phase gain of plane waves propagating through the {FP} cavity.
By using the equation of motions for Green function and the Fermi liquid theory~\cite{Pustilnik04, Hewson93}, $S_\mathrm{QD}$ and $S_\mathrm{QPC}$ are obtained as
\begin{align}
S_\mathrm{QD} &= \frac{1}{t_\mathrm{l}^2 + t_\mathrm{r}^2}\left[
\begin{array}{cc}
e^{2i\delta}t_\mathrm{l}^2 + t_\mathrm{r}^2 & (e^{2i\delta}-1)t_\mathrm{1} t_\mathrm{r}
\\
(e^{2i\delta}-1)t_\mathrm{l}t_\mathrm{r} & t_\mathrm{l}^2 + e^{2i\delta}t_\mathrm{r}^2
\end{array}
\right], \label{S_QD} \\
S_\mathrm{QPC} &=
\frac{1}{2-\alpha}
\left[
\begin{array}{cc}
\alpha & 2i\sqrt{1-\alpha}
\\
2i\sqrt{1-\alpha} & \alpha
\end{array}
\right]. \label{S_QPC}
\end{align}
Here $\delta$ is the scattering phase shift by the QD, and $\delta = \pi/2$ at zero temperature in the Kondo regime. $S_\mathrm{tot}$ can be obtained from
\begin{align}
M_\mathrm{tot}
= M_\mathrm{QPC}
\left[
\begin{array}{cc}
e^{ikL} & 
\\
 & e^{-ikL}
\end{array}
\right]
M_\mathrm{QD}.
\end{align}
Here $e^{ikL}$ is the dynamical phase gain of an electron with momentum $k$ propagating through the FP cavity, and the matrices $M$'s are the transfer matrices related with the corresponding scattering matrix via
\begin{align}
S_a = \left[
\begin{array}{cc}
s_{11} & s_{12}
\\
s_{21} & s_{22}
\end{array}
\right]
\leftrightarrow
M_a = \frac{1}{s_{22}}
\left[
\begin{array}{cc}
s_{12}s_{21} - s_{11}s_{22} & s_{22}
\\
-s_{11} & 1
\end{array}
\right],
\end{align}
for $a= \mathrm{tot}, \mathrm{QD}, \mathrm{QPC}$.
We find that the transmission amplitude $t_\mathrm{tot}$ through the whole scattering region containing the QD and the FP cavity is written as
\begin{align}\label{T_tot}
t_\mathrm{tot} = \frac{-2i\sqrt{1-\alpha}(e^{2i\delta}-1)t_\mathrm{l} t_\mathrm{r}e^{i k_\mathrm{F} L}}
{[2-(1+e^{2i k_\mathrm{F} L})\alpha]t_\mathrm{l}^2 + [2-(1+e^{2i\delta}e^{2ik_\mathrm{F} L})\alpha]t_\mathrm{r}^2}
\end{align}
for electrons at the Fermi level.
 
\subsection{The Kondo regime}\label{t21}

We apply the scattering-theory result in Eq.~\eqref{T_tot} to the case that the QD is in the Kondo regime. In this case, the scattering phase shift $\delta = \pi/2$ at zero temperature, according to the Fermi liquid theory. Using the estimated parameters in Table~\ref{par}, we compute the zero-temperature conductance $G_0 = 2e^2 | t_\mathrm{tot}|^2/h$.
As shown in Fig.~\ref{fig_t1}, the conductance $G_0$ obtained by using the scattering theory is in good agreement with the conductance obtained by the NRG formula in Eq.~\eqref{G}. Interestingly, the conductance exhibits the maximum value at the off-resonance conditions of $e^{2ik_\mathrm{F}L} = -1$, and shows the minimum value at the on-resonance conditions of $e^{2ik_\mathrm{F}L} = 1$. This results from the phase shift $\delta = \pi/2$. On the other hand, the Kondo temperature has the maximum (minimum) value at the on-resonance (off-resonance) conditions; note that the same conclusion is obtained when the Kondo temperature is estimated by using $T_\mathrm{K} = \exp (- \pi U / 8 \Gamma)$ and $\Gamma = \Gamma_\mathrm{l}(E=\epsilon_\mathrm{F})+\Gamma_\mathrm{r}(E=\epsilon_\mathrm{F})$. 
Therefore, the oscillations of $G_0$ and $T_\mathrm{K}$ as a function of $2 k_\mathrm{F} L$ are out of phase by $\pi$. This out-of-phase behavior occurs in both the regimes of $L / \xi_{\mathrm{K} \infty} > 1$ and $L / \xi_{\mathrm{K} \infty} < 1$, and at low temperature $T \lesssim T_\mathrm{K}, \Delta$.
This agrees with the experimental data in Fig.~2.

\subsection{A single noninteracting level}\label{t22}

We note that the out-of-phase behavior of the oscillations of $G_0$ and $T_\mathrm{K}$ occurs also in the case that the QD effectively has only a single noninteracting level at the Fermi level. It is because the QD energy level aligned with the Fermi level results in the scattering phase shift $\delta = \pi/2$ as in the Kondo regime.
This argument is applicable to the case that the QD shows Coulomb blockade resonance, and it agrees with our experimental data.

\section{Validity of estimating $T_\mathrm{K}$'s}\label{t30}

\subsection{Fitting of the experimental data}
\begin{figure}[h]
\includegraphics[width=0.5 \columnwidth]{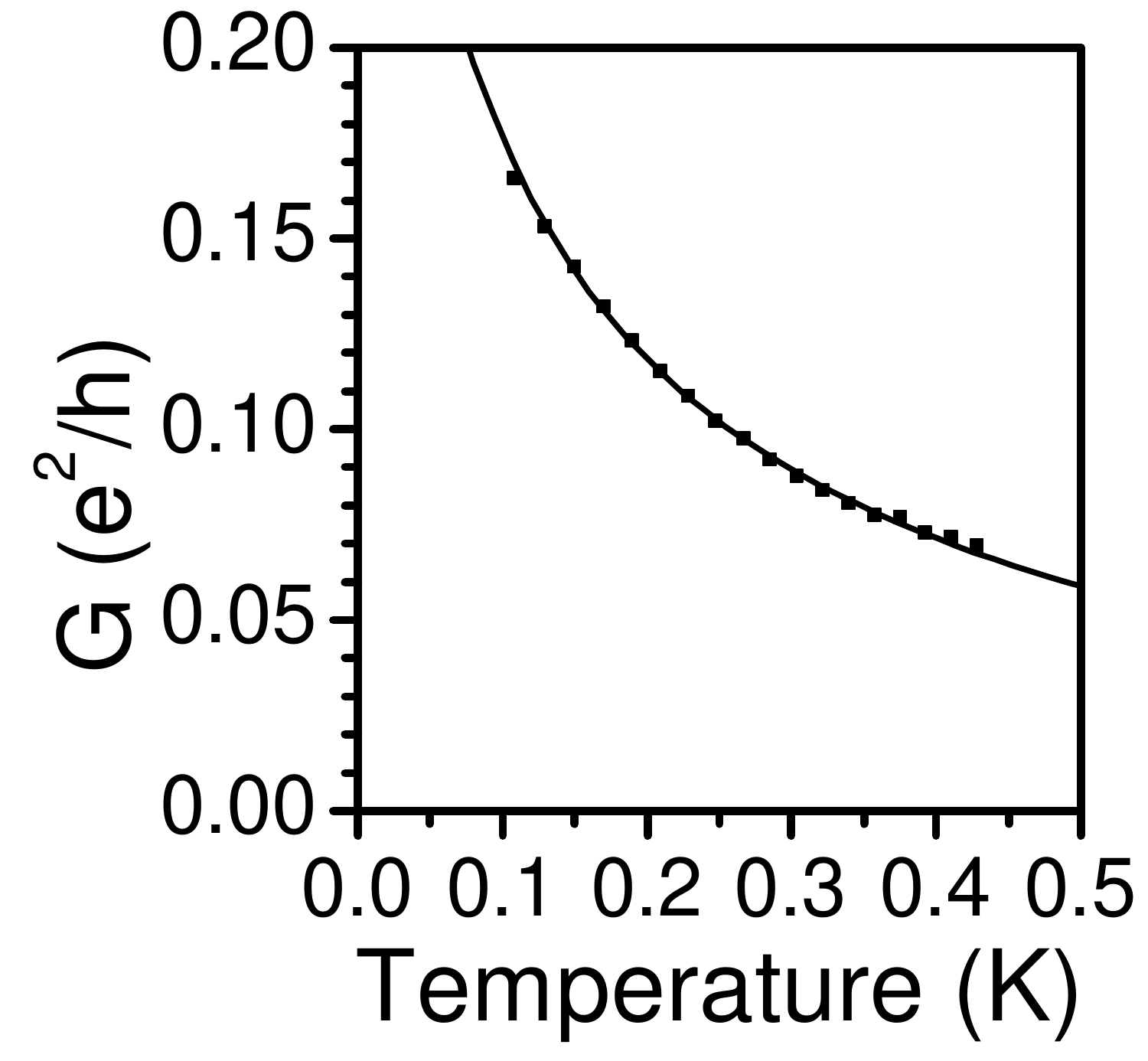}
\caption{\label{fig_e_2} Plot of conductance versus temperature of the center of the Kondo valley. Dashed lines present a fit to the empirical formula. 
}
\end{figure}

The Kondo temperature $T_\mathrm{K}$ is estimated by measuring the center of the Kondo valley conductance $G$ as a function of temperature $T$. The trend of $G$ vs $T$ is described well by the empirical formula in Eq.~\eqref{empiricalF} with $s=0.22 \pm 0.01$ \cite{Empirical_Kondo}. An example of the fit of our data to the empirical formula is shown in Figure~\ref{fig_e_2}, with $T_\mathrm{K}$ and $G_0$ as fitting parameters. 
Our NRG calculations support the ways of estimating $T_\mathrm{K}$ and $T_{\mathrm{K} \infty}$ from our experimental data, as shown below.

\subsection{Compatibility with NRG calculations: Estimating $T_\mathrm{K}$ }\label{t32}
The above estimation works well when the hybridization functions of the reservoirs coupled to the QD are energy independent near the Fermi level over the energy scale much larger than Kondo temperature.
In our experiment, however, the hybridization function $\Gamma_\mathrm{r}$ of the right wire is energy dependent, because of the resonances formed in the {FP} cavity.
Below we show, based on the NRG calculation, that the empirical formula is still useful for estimating $T_\mathrm{K}$ in our experiment regime of $L \lesssim \xi_{\mathrm{K}\infty}$.

\begin{figure}[t]
\centerline{\includegraphics[width=0.5\textwidth]{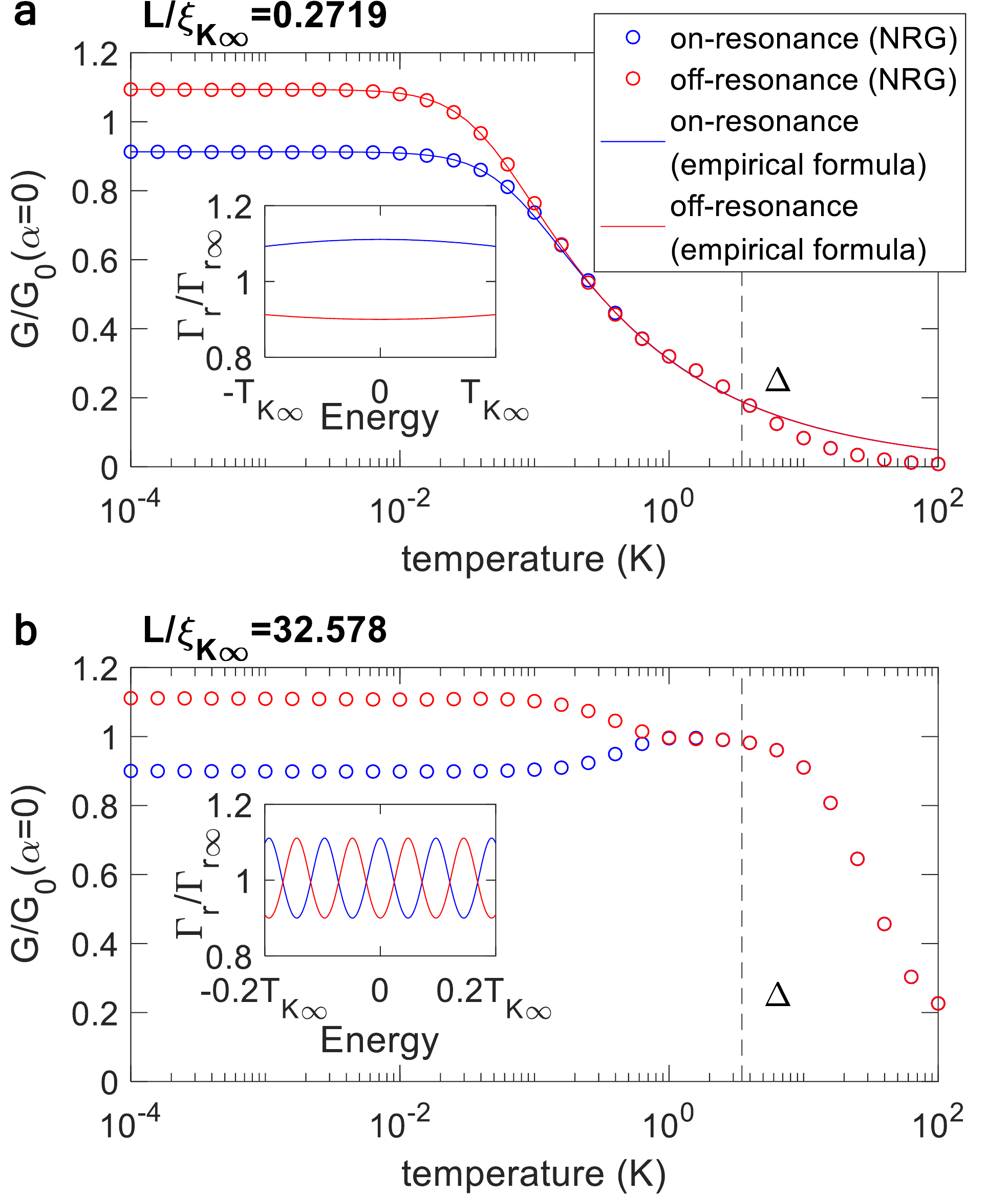}}
\caption{NRG results of the conductance $G$ as a function of temperature $T$ at on-resonance (blue circles) and off-resonance (red) in the cases of (a) $L < \xi_{\mathrm{K}\infty}$ and (b) $L > \xi_{\mathrm{K}\infty}$. In (a), the NRG results in the temperature window of [0 K, 0.5 K] are well fitted to the empirical formula (see solid curves). 
The vertical dashed lines indicate the temperature comparable to the resonance level spacing $\Delta$.
Inset: The hybridization function $\Gamma_\mathrm{r}$ as a function of energy $E$ at on-resonance (blue curve) and off-resonance (red).
}
\label{fig_t3}
\end{figure}

Figure~\ref{fig_t3}a shows the NRG result of the temperature dependence of the conductance in a case of $L \lesssim \xi_{\mathrm{K}\infty}$. In this case, the resonance level spacing $\Delta$ of the {FP} cavity is larger than Kondo temperature $T_{\mathrm{K}\infty}$, hence, the hybridization function $\Gamma_r$ is almost energy independent near the Fermi level (see the inset of Fig.~\ref{fig_t3}a) both in the situations of on- and off-resonance; the zero-temperature conductance at on-resonance is smaller than that at off-resonance due to the phase shift $\delta = \pi/2$ as discussed in Sec.~\ref{t21}. We notice that the NRG result is well fitted to the empirical formula in this case, although the NRG result slightly deviate from the empirical formula around the temperature comparable to the resonance level spacing $\Delta$.

The deviation becomes pronounced in the opposite regime of $L \gg \xi_{\mathrm{K}\infty}$ (note that our experiment is not in this regime). In this regime, there are many resonances within the energy window of $T_{\mathrm{K}\infty}$ near the Fermi level, as shown in the inset of Fig.~\ref{fig_t3}b. As a result, the conductance has unusual temperature dependence at temperature comparable to the resonance level spacing $\Delta$, hence, the empirical formula does not well describe the temperature dependence. For example, at on-resonance, the conductance is almost constant near $T=0$, increases slowly with temperature at $T \lesssim \Delta$, and then decreases at $T \gtrsim \Delta$. This is because the conductance is governed by a single resonance near $T=0$, and then by multiple resonances near $T \sim \Delta$.

We note that similar but stronger deviation can occur in the regime of a Kondo box~\cite{Kondobox1,Kondobox2,Kondobox3}.
 
\subsection{Compatibility with NRG calculations: Estimating $T_{\mathrm{K} \infty}$}\label{t31}

In the experiment, $T_{\mathrm{K}\infty}$, the Kondo temperature of the case where no QPC is formed in the right wire, cannot be directly obtained, since the QPC gate affects the right wire even when the gate voltage  $V_\mathrm{QPC}$ is turned off. We estimate it as $T_{\mathrm{K}\infty} \approx \sqrt{T_\mathrm{K,Max} T_\mathrm{K,Min}}$, the geometric mean of $T_\mathrm{K,Max}$ and $T_\mathrm{K,Min}$, where $T_\mathrm{K,Max}$ ($T_\mathrm{K,Min}$) is the maximum (minimum) value of $T_\mathrm{K}$ among those obtained with tuning $2k_\mathrm{F} L$ in the experiment.

\begin{figure}[h]
\centerline{\includegraphics[width=0.75\textwidth]{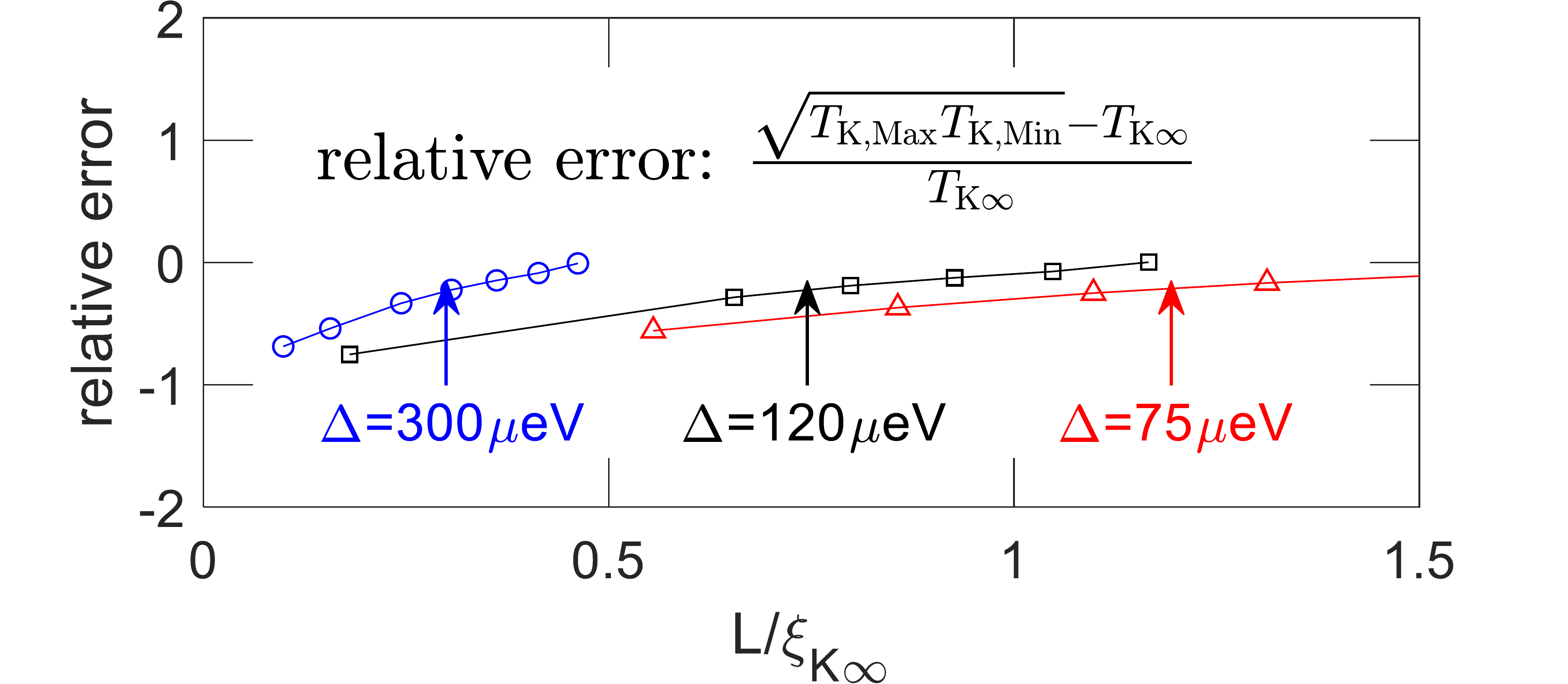}}
\caption{Comparison between $T_{\mathrm{K}\infty}$ and $ \sqrt{T_\mathrm{K,Max}T_\mathrm{K,Min}}$, which are obtained by using the NRG method.
}
\label{fig_t2}
\end{figure}

In Fig.~\ref{fig_t2} we compare $T_{\mathrm{K}\infty}$ and $\sqrt{T_\mathrm{K,Max} T_\mathrm{K,Min}}$, which we compute by using the NRG, the Hamiltonian in Eq.~\eqref{modelH} and the parameters in Table~\ref{par}. 
This shows that the estimation of $T_{\mathrm{K}\infty}$ from the experimental data of $\sqrt{T_\mathrm{K,Max} T_\mathrm{K,Min}}$ is valid.

We expect such good agreement between $T_{\mathrm{K}\infty}$ and $\sqrt{T_\mathrm{K,Max} T_\mathrm{K,Min}}$ in our experimental regime of $t_\mathrm{l} \ll t_\mathrm{r}$, $\Delta \gtrsim T_{\mathrm{K}\infty}$, and small $\alpha$ satisfying $(1-\alpha)^{-1} \approx 1+\alpha$.
In this regime, we find $T_\mathrm{K,Max} \approx (T_{\mathrm{K}\infty})^{1-\alpha}$ and $T_\mathrm{K,Min} \approx (T_{\mathrm{K}\infty})^{1+\alpha}$, by using
the expression~\cite{Hewson93} of $T_{\mathrm{K}\infty} \sim \text{exp}(-\frac{\pi U}{8\Gamma_{r\infty}(E=\epsilon_\mathrm{F})})$ at the on- and off-resonances of  $e^{2ik_\mathrm{F}L}= \pm 1$. This provides $T_{\mathrm{K}\infty} \approx \sqrt{T_\mathrm{K,Max} T_\mathrm{K,Min}}$.

\section{Universal scaling at $L < \xi_{\mathrm{K}\infty}$}\label{t40}

The experimental data in Fig.~3 of the main text show the behavior of 
\begin{equation} \text{ln}(T_\mathrm{K,Max}/T_{\mathrm{K}\infty}) \approx -\eta\text{ln}(L / \xi_{\mathrm{K}\infty}) \label{UniScaling} \end{equation} 
at $L < \xi_{\mathrm{K}\infty}$.
This behavior is scaled only by $T_{\mathrm{K}\infty}$ and $\xi_{\mathrm{K}\infty}$, hence, it is a universal feature characterizing the core region ($\hbar v_\mathrm{F} / U < L < \xi_{\mathrm{K}\infty}$) of the Kondo cloud. 

We derive the behavior. In our experimental regime of $t_\mathrm{l} \ll t_\mathrm{r}$ and small $\alpha$, the poorman scaling~\cite{Park13} leads to $\text{ln}(T_\mathrm{K,Max}/T_{\mathrm{K}\infty}) \approx -\alpha \text{ln}(L / \xi_{\mathrm{K}\infty})$. In this estimation based on the poorman scalie, $\eta$ equals $\alpha$.This implies that the coefficient $\eta$ is determined mainly by the resonance broadening parameter $\alpha$ for small $\alpha$.
The NRG calculation confirms the universal scaling in Eq.~\eqref{UniScaling}, but with $\eta \ne \alpha$. The value of $\eta$ depends on the estimation method of the Kondo temperature; the Kondo temperature is defined in the poorman scaling in a way different from that in the NRG method.

We find that the scaling behavior also occurs in the other regime of $t_\mathrm{l,r}$ and $\alpha$ including the regimes of large $\alpha$.

We also note that at $L > \xi_{\mathrm{K}\infty}$, the Kondo cloud has a long tail following another algebraic scaling law characterized by quantum entanglement or electron conductance~\cite{KondoEntanglement_NRG,KondoEntanglement_conductance}.

\end{document}